# Multi-MeV electron acceleration by sub-terawatt laser pulses

A.J. Goers, G.A. Hine, L. Feder, B. Miao, F. Salehi, and H.M. Milchberg

*Institute for Research in Electronics and Applied Physics*
*University of Maryland, College Park, MD 20742*

We demonstrate laser-plasma acceleration of high charge electron beams to the ~10 MeV scale using ultrashort laser pulses with as little energy as 10 mJ. This result is made possible by an extremely dense and thin hydrogen gas jet. Total charge up to ~0.5 nC is measured for energies >1 MeV. Acceleration is correlated to the presence of a relativistically self-focused laser filament accompanied by an intense coherent broadband light flash, associated with wavebreaking, which can radiate more than ~3% of the laser energy in a sub-femtosecond bandwidth consistent with half-cycle optical emission. Our results enable truly portable applications of laser-driven acceleration, such as low dose radiography, ultrafast probing of matter, and isotope production.

Laser-driven electron acceleration in plasmas has achieved many successes in recent years, including record acceleration up to 4 GeV in a low emittance quasi-monoenergetic bunch [1] and generation of high energy photons [2-5]. In these experiments, the driver laser pulse typically propagates in the 'bubble' or 'blow-out' regime [6, 7] for a normalized peak vector potential $a_0 = eA_0/mc^2 \gg 1$. Plasma densities are deliberately kept low for resonant laser excitation and to avoid dephasing [7]. Essentially all of these experiments use 10 TW − 1 PW laser drivers, with repetition rates ranging from 10 Hz to an hour between shots [8].

For many modest lab scale and portable applications, however, a compact, relatively inexpensive, high average current source of laser-accelerated relativistic electrons is sufficient and desirable. In this paper we describe experiments using a very dense, thin hydrogen gas jet, where the relativistic self-focusing threshold is exceeded even with ~10 mJ laser pulses and MeV-scale energy electron bunches are generated. This enables applications, such as ultrafast low dose medical radiography, which would benefit from a truly portable source of relativistic particle beams. We note that prior work has shown electron bunch generation of modest charge and acceleration (~10 fC/pulse, <150 keV) from a 1 kHz, ~10 mJ laser driving a thin (~100 μm), low density continuous flow argon or helium jet [9].

Central to our experiment is a thin, high density pulsed hydrogen sonic gas jet, which reaches a maximum peak molecular density of $9\times10^{20}$ cm$^{-3}$, and when fully ionized can exceed the plasma critical density, $N_{cr}=1.7\times10^{21}$ cm$^{-3}$ at our laser wavelength of $\lambda_0$=800nm. The density profile is near-Gaussian, with a full width at half maximum (FWHM) in the range 150-250 μm, depending on the height of the optical axis above the jet orifice. Earlier versions of this jet were run in both pulsed [10] and continuous flow [11] for nitrogen and argon. High densities are achieved using a combination of high valve backing pressure and cryogenic cooling of the valve feed gas, which is forced through a 100μm diameter needle orifice. Cooling to −160C enables a significant density increase for a given valve backing pressure. Figure 1 shows the experimental setup. Pulses from a Ti:Sapphire laser (50 fs, 10-50 mJ) are focused into the gas jet (a) at f/9.5 by a 15° off-axis parabolic mirror. Inset (b) shows neutral hydrogen profiles measured by interferometry. A wavefront sensor and deformable mirror are used to optimize the vacuum focal spot FWHM to $w_{FWHM}$ =8.4 µm, or 1.2× the diffraction limit, with a confocal parameter of 550 µm and 80% of the pulse energy in the near-Gaussian spot. For maximizing electron beam charge and energy, it was found that placing the focused beam waist at the centre of the gas jet was optimal without strong sensitivity to positioning. This is consistent with the laser confocal parameter being more than twice the jet width. We note that recent laser interaction experiments at near critical density [12] have used a complex pressure-boosted mm-scale gas jet [13].

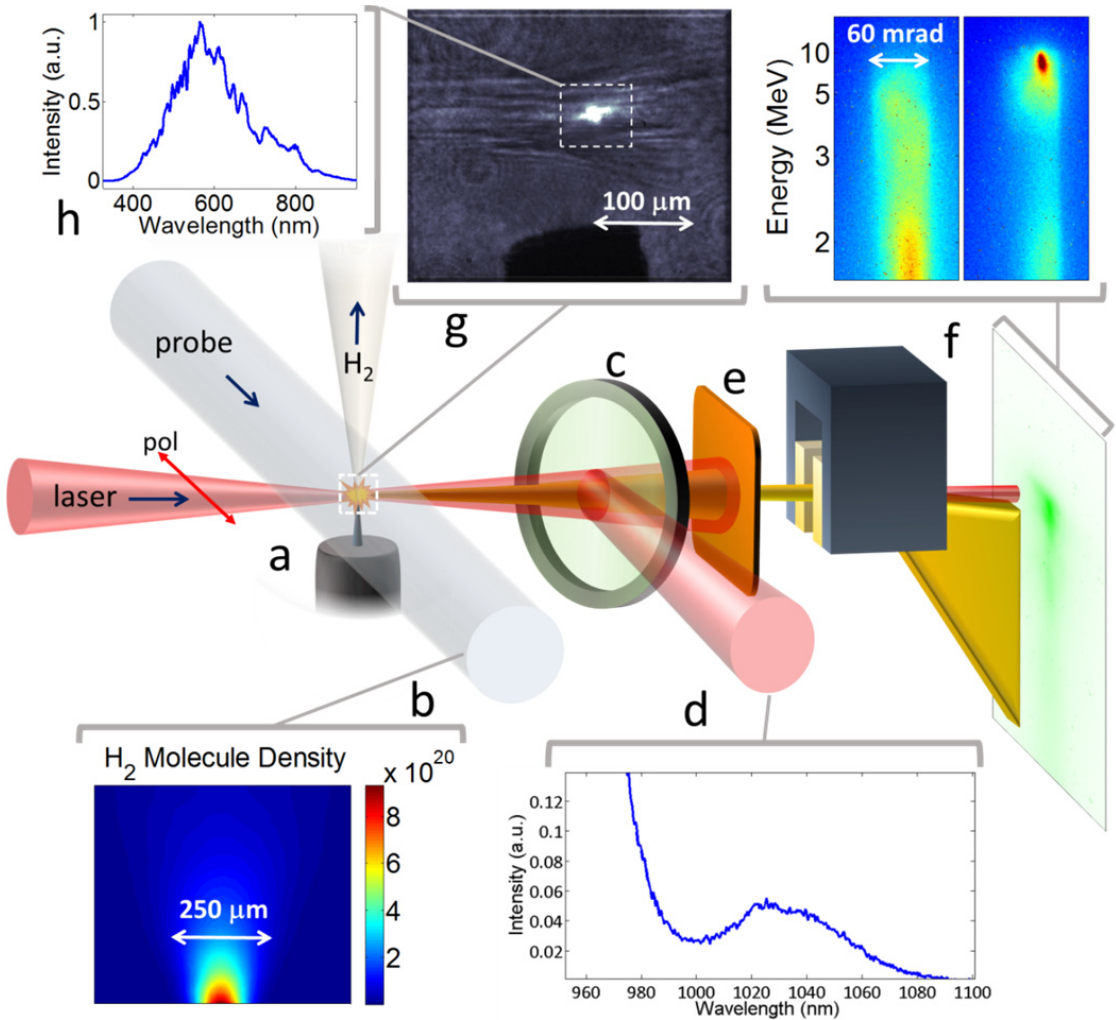

**Figure 1.** Experimental setup. A horizontally polarized Ti:Sapphire laser pulse (10-50 mJ, 50 fs, λ=800 nm) interacts with a cryogenically-cooled, ultra-dense thin $H_2$ gas jet (a), whose neutral and plasma density profiles are measured by 400 nm probe interferometry (b). A portion of the transmitted laser pulse is reflected by a pellicle (c) and measured by a spectrometer (d). The electron beam from the jet is apertured by a 1.7mm horizontal slit (e), enters a 0.13 T permanent magnet spectrometer, and is dispersed on an aluminum foil-shielded LANEX screen (f), which is imaged by a low noise CCD camera (not shown). (f) shows example quasi-monoenergetic and exponential spectra for a 40 mJ pulse at $N_e$=2×10$^{20}$ cm$^{-3}$. Shadowgraphic imaging of the laser interaction region above the needle orifice (g) (needle seen as a shadow at bottom) and imaging (g) and spectroscopy (h) of the wavebreaking flash. The pump polarization could also be rotated to the vertical by a half wave plate.

The neutral jet density and plasma profiles were measured using a 400nm, 70fs probe pulse (derived from the main pulse), which was directed perpendicularly through the gas jet to a folded wavefront interferometer. Forward- and side-directed optical spectra were collected by fibre-coupled spectrometers, with the forward spectra directed out of the path of the pump laser and electron beam by a pellicle. Shadowgraphic images using the 400 nm probe and images of bright broadband wavebreaking radiation flashes were collected using achromatic optics.

Relativistic electron spectra in the energy range 2–15 MeV were measured using a 0.13 T permanent magnet spectrometer 25 cm downstream of the gas jet. A copper plate with a 1.7 mm × 12 mm slit aperture in front of the magnet entrance provided energy resolution while allowing measurement of beam divergence in 1D. Electron spectra were dispersed along a LANEX scintillating screen, shielded against exposure to the laser by 100 μm thick aluminum foil, and imaged using a low noise CCD camera. Full electron beam profiles were collected by the LANEX screen by translating the dispersing magnets and slit aperture out of the way. Estimates of the accelerated charge were made by calibrating the imaging system and using published LANEX conversion efficiencies [14].

The high density of our target has the immediate effect of enabling relativistic self-focusing of low energy laser pulses and generation of a nonlinear plasma wake. Furthermore, the reduced laser group velocity (and therefore plasma wave phase velocity) at high density drops the threshold for electron injection. Figure 2 shows > 1 MeV electron beam generation for pulse energies in the range 10-50 mJ, or 0.2–1.0 TW, as a function of peak plasma density. Beam divergence is ≲ 200 mrad. The results are consistent with the inverse density scaling of the self-focusing critical power, $P_{cr} = 17.4(N_{cr}/N_e)$ GW [15], and the laser power threshold for appearance of a relativistic electron beam is ~ $3P_{cr}$ across our range of conditions.

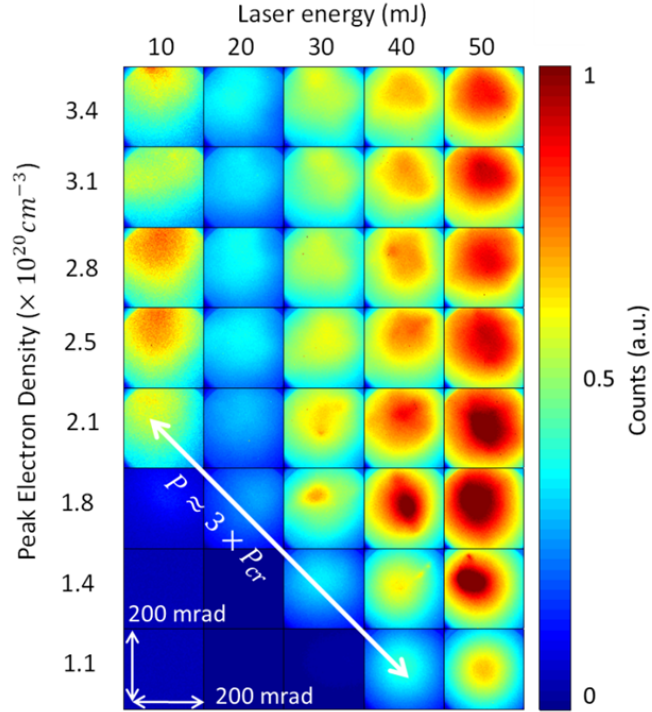

**Figure 2.** Single shot electron beam images for energies > 1 MeV for a range of laser energies and peak profile electron densities. The colour palette was scaled up by 10× for the 10 mJ column. The onset laser power for detectable electron beam generation was ~$3P_{cr}$ across our range of conditions.

Electron energy spectra in the range 2–12 MeV are shown in Fig 3(a) for laser pulse energy 10-50 mJ and peak electron density $N_e$=4.2×10$^{20}$ cm$^{-3}$, with the inset showing total accelerated charge > 2 MeV up to ~1.2 nC/sr for 50 mJ laser pulses. An electron spectrum simulated from a TurboWAVE 3D particle in cell (PIC) simulation [16] for the 40 mJ case is overlaid on the plot. Electron spectra as a function of peak density for fixed pulse energy of 40 mJ are shown in Fig. 3(b) along with results from the 3D PIC simulations. We note that for approximately 20% of shots near the self-focusing onset at each pressure we observed quasi-monoenergetic peaks ranging from 3 MeV (~25 fC for 10 mJ) to 10 MeV (~1.4 pC for 50 mJ, see Fig. 1(f)) with ~10 mrad beam divergence. Both the spectra and the beam spot positions are highly variable and are the subject of ongoing work.

Another consequence of the high density gas target is that the pump pulse envelope is multiple plasma periods long. Over our experimental density range of $N_e$=1-4×10$^{20}$ cm$^{-3}$, the plasma period is $2\pi/\omega_p$ ~ 11fs–5.7 fs, placing our 50 fs pump pulse in the self-modulated laser wakefield acceleration (SM-LWFA) regime. Evidence of SM-LWFA is seen in the moderately collimated electron beams of Fig. 2 and the exponential electron spectra of Fig. 3, reflecting acceleration from strongly curved plasma wave buckets and electron injection into a range of accelerating phases. This is consistent with prior SM-LWFA experiments [17], except that here our dense hydrogen jet enables production of MeV spectra with laser pulses well below 1 TW. Further confirmation of self-modulation is shown in the sample spectrum of Raman forward scattered Stokes radiation shown in the inset of Fig. 1, for the case of laser energy 50 mJ (vacuum $a_0 \sim 0.8$) and peak density $N_e$ =1.8×10$^{20}$ cm$^{-3}$. The strong broadband red-shifted Raman peak located at $\lambda_s = 2\pi c/\omega_s$ ~ 1030 nm enables the estimate of self-focused $a_0$~2.7, using the measured electron density profile and $\omega_s = \omega - \omega_p/\gamma^{1/2}$, where $\omega$ is the laser frequency and $\gamma = (1 + a_0^2/4)^{1/2}$ is the relativistic factor. This estimate is in good agreement with the peak $a_0$ in our 3D PIC simulations.

In order for electrons to be accelerated, they must first be injected into the wakefield. Our 3D simulations show catastrophic transverse wavebreaking [18] of the strongly curved plasma wavefronts [19] behind the laser pulse, which injects electrons from a wide spread of initial trajectories into a range of phases of the plasma wave. Wavebreaking is accompanied by an extremely strong broadband radiation flash emitted by electrons accelerated from rest to near the speed of light in a small fraction of a plasma wavelength. Figure 1 shows a magnified single shot image of the sideways-collected flash superimposed on a shadowgram image of the relativistically self-focused filament, while Fig. 4 shows 10 shot average images of the flash for varying plasma peak density and laser energy, for horizontal pump polarization. Such radiation has been observed in prior work, although at a much lower energy and yield (~0.1 nJ for a 500 mJ pump pulse) [20]. Here neutral density filters were employed to prevent the side-imaged flash intensity from saturating our CCD. We measure flash energies of ~15 μJ into f/2.6 collection optics for the 40 mJ, $N_e$ =3.4×10$^{20}$ cm$^{-3}$ panel in Fig. 4, giving ~1.5 mJ or >3% of the laser energy into 4π sr. The flash energy, spectrum and axial location are independent of pump polarization and thus the flash does not originate from pump scattering. Broadband flash spectra (10 shot averages, with no filtering of the pump), peaking at $\lambda_{rad}$~550-600nm with bandwidth ~400nm, are shown at the bottom of Fig. 4 for 40 mJ and a range of densities. The figure panels show that the flash occurs on the hydrogen density profile up-ramp for higher densities and laser energies and on the down-ramp for lower densities and laser energies, as also borne out by our 3D simulations. This is explained by the earlier onset of relativistic self-focusing for either higher density (for fixed laser energy) or higher laser energy (for fixed density), which is followed closely by self-modulation and wavebreaking.

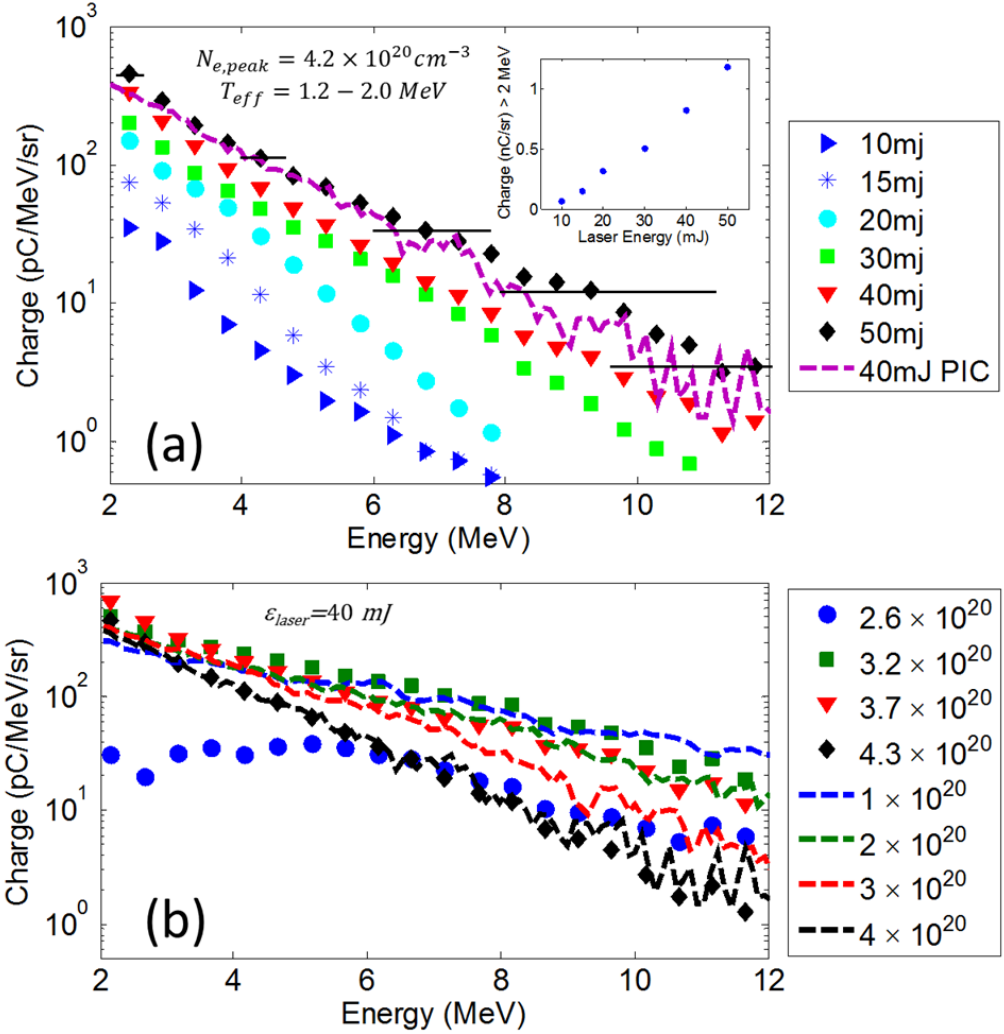

**Figure 3.** (a) Accelerated electron spectra for peak jet electron density $4.2 \times 10^{20}$ cm$^{-3}$ for varying laser energy. The inset shows total charge >2 MeV as a function of laser energy. The range of effective temperatures of these exponential-like distributions is indicated. The horizontal black lines indicate the experimental uncertainty in the energy, determined by geometry-limited spectrometer resolution. The dashed curve is a 3D PIC simulation for 40 mJ pump which has been scaled by a factor 0.14 to line up with the experimental curve for 40 mJ. (b) Accelerated electron spectra at laser energy 40 mJ for varying peak electron density. The dashed curves are from 3D PIC simulations and were scaled by the factor 0.14.

The huge increase in radiation flash energy compared to earlier experiments [20] stems from its coherent emission by electron bunches wavebreaking over a spatial scale much smaller than the radiation wavelength and the consequent damping of these bunches by this radiation. As a rough estimate of this effect in 1D, the near-wavebreaking crest width $\Delta x_{crest}$ of the nonlinear plasma wave is given by $\Delta x_{crest} / \lambda_p \sim \frac{1}{\pi}(\omega/\omega_p)^{3/4}(\Delta p_0 / 2mc)^{3/4}$ [21], where $\Delta p_0$ is the electron initial longitudinal momentum spread. For $N_e = 3 \times 10^{20}$ cm$^{-3}$ and $\Delta p_0 / mc \sim 0.06$ (from an initial spread $\sim (\Delta p_0)^2 / 2m$ < 1 keV from residual electron heating after ionization [22]) we get $\Delta x_{crest} / \lambda_p \sim 0.04$, or $\Delta x_{crest} \sim 0.12 \lambda_{rad}$, significantly shorter than the peak radiated wavelength, ensuring coherent emission. The wavebreaking electrons radiate as they execute curved orbits into the wake bucket just ahead of the crest, with emission power $P \sim n^2 (\frac{2e^2}{3m^2c^3} \gamma^2 \left|\frac{d\mathbf{p}}{dt}\right|^2)$ [23] and total radiated energy $\varepsilon_{rad} \sim P\Delta t \sim P\Delta x_{crest} \gamma^{-1} / c$, where $n$ is the number of accelerating electrons from the crest, $d\mathbf{p}/dt$ is the force on an electron, and $\Delta t$ is the sub-femtosecond time for acceleration off the crest (accounting for Lorentz contraction), or equivalently, the crest lifetime. Taking $n \sim N_e \lambda_p^3$, $\gamma = \gamma_p = \omega/\omega_p$, and $|d\mathbf{p}/dt| \sim eE_{wb}$, where $E_{wb} = \sqrt{2}\,(m\omega_p c/e)(\gamma_p - 1)^{1/2}$ is the 1D wavebreaking field [18], and using $N_e = 3\times 10^{20}$ cm$^{-3}$, gives a total radiated energy of $\varepsilon_{rad} \sim 2$ mJ, which is of the correct order of magnitude.

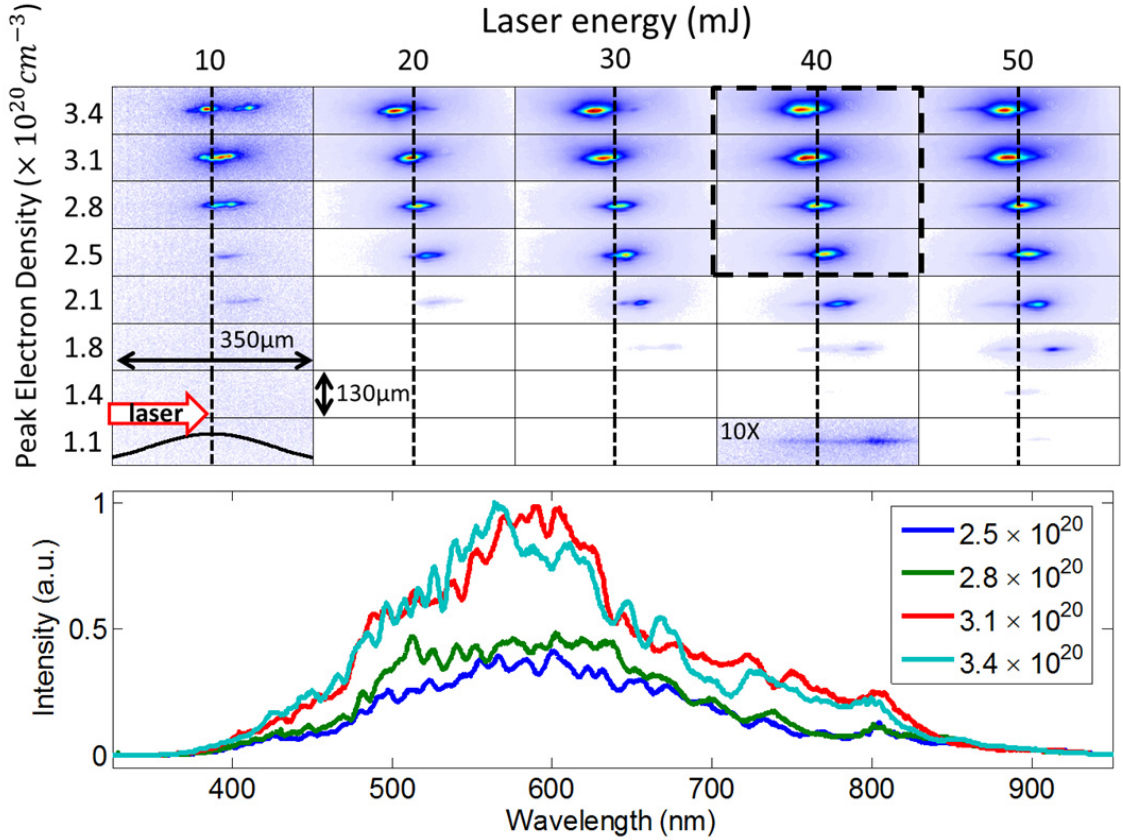

**Figure 4.** *Top panel:* Side images of intense radiation flashes from wavebreaking (10 shot averages). The pump laser pulse propagates left to right. Image intensities are normalized to the maximum intensity within each column. The vertical dashed line is the centre of the gas jet, whose profile is shown in the lower left. The 40 mJ, $1.1\times 10^{20}$cm$^{-3}$ image for vertical pump polarization (enhanced 10×), is dominated by 800 nm Thomson scattering on the left and the flash on the right. *Bottom panel:* Spectra (10 shot averages) of the flash for conditions enclosed by the dashed black box in the top panel.

The strong curving of injected electron orbits by the ion column allows an estimate of the flash spectrum range using the synchrotron radiation critical frequency $\omega_{rad} \sim \omega_c \sim 3c\gamma^3/2\rho$ [23], where $\rho$ is the orbit radius of curvature, which we take as $\rho \sim \lambda_p$, assuming trapping within a single plasma wave bucket. This gives $\lambda_c \sim 4\pi/3(N_e/N_{cr})\lambda_0 \sim$ 400-700nm for the density range shown in Fig. 4, reasonably overlapping the measured spectra. The synchrotron spectrum bandwidth is estimated as $(\Delta\lambda/\lambda)_{rad} \sim (\omega_{rad}\Delta t)^{-1} \sim 1$, in accord with the sub-femtosecond bandwidth in Fig. 4 characteristic of half-cycle optical emission, which is consistent with the violent unidirectional electron acceleration upon wave breaking. To our knowledge, this is the first evidence of bright half-cycle optical emission. We note that half-cycle wavebreaking radiation at high $\gamma$ has been proposed as an attosecond x-ray source [24].

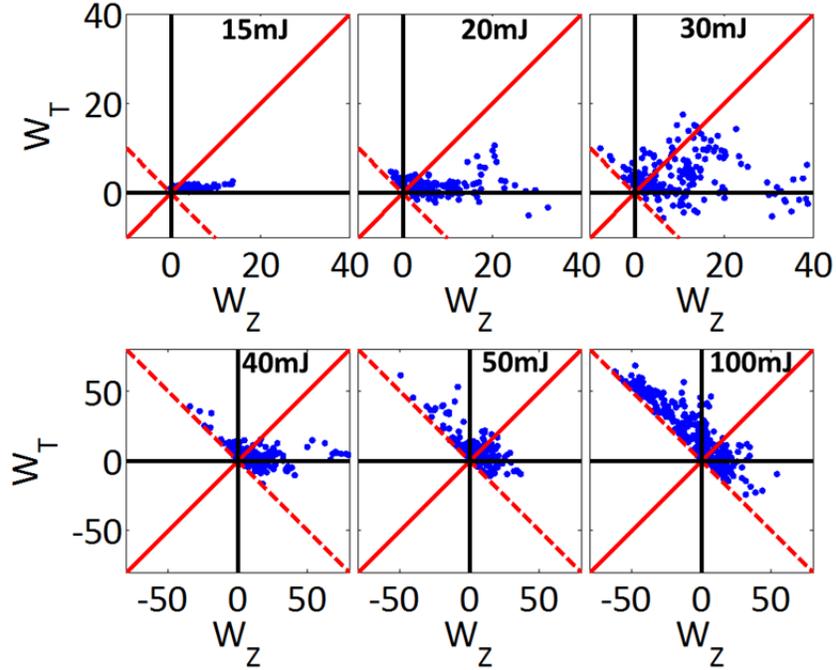

**Figure 5.** 2D PIC simulations showing contributions of LWFA and DLA to electron energy gain for a fixed peak plasma density $N_e = 0.07 N_{cr}$ for drive laser energies 15-100mJ. Each blue dot is a tracked electron. Regions above and to the left of the solid red line indicate DLA as the dominant form of acceleration, whereas regions below and to the right are dominated by LWFA. The dashed red diagonal marks zero net energy gain. LWFA dominates acceleration at low drive laser energies, transitioning to DLA at high drive laser energies.

A question that has arisen in prior experiments and simulations of acceleration at higher plasma densities [25-27] is the relative contributions of laser wakefield acceleration (LWFA) and direct laser acceleration (DLA). The contribution of the plasma wave to electron energy gain (in units of $mc^2$) is $W_z = -\frac{e}{mc^2}\int E_z v_z dt$, where the integral is over the full electron trajectory and $E_z$ and $v_z$ are the longitudinal plasma wakefield and electron velocity. The DLA contribution, $W_\perp = -\frac{e}{mc^2}\int \mathbf{E}_\perp \cdot \mathbf{v}_\perp dt$, arises as near light-speed electrons, axially co-propagating with the laser field $\mathbf{E}_\perp$, oscillate with a $\mathbf{v}_\perp$ component about the ion column axis at the betatron frequency $\omega_\beta = \omega_p/\sqrt{2\gamma}$ or its harmonics, in resonance with the field [25]. In our experiment, the jet location-dependence of electron injection determines the relative contributions of DLA and LWFA to the net energy gain. Early wavebreaking injection on the density profile up-ramp occurs when the plasma wavelength is decreasing and more wake buckets lie under the laser pulse envelope, exposing injected electrons to DLA. On the down-ramp, the plasma wavelength is increasing and fewer buckets lie under the laser field envelope, so that injected electrons are less exposed to the laser field. The flash images of Fig. 4 are a map of injection locations through the jet, therefore they spatially map the relative balance of DLA and LWFA, predicting that DLA dominates at high density /high laser energy and LWFA dominates at low density /low laser energy. This transition from LWFA to DLA is corroborated by 2D PIC simulations. Figure 5 shows the contributions for a subset of tracked particles which were accelerated to energies above 1 MeV. For a fixed peak density $N_e = 0.07 N_{cr}$, the figure panels show clearly that acceleration with low laser energies is dominated by LWFA, transitioning to DLA at energies above ~50 mJ.

In summary, we have demonstrated electron acceleration to the 10 MeV scale with laser pulses well below 1 terawatt, using a thin, high density hydrogen gas jet, with efficiency of laser energy to MeV electrons of a few percent. The high plasma density reduces the thresholds for relativistic self-focusing, nonlinear plasma wave generation, and electron injection. The reduced spatial scales associated with high density yields intense coherent wavebreaking radiation, whose sub-femtosecond bandwidth is consistent with half-cycle optical emission upon violent unidirectional electron acceleration from rest to nearly the speed of light over a subwavelength distance. The flash is of sufficient intensity to self-damp the injected bunch, with the result that wavebreaking radiation and acceleration are comparable energy channels. Our results open the way to applications of relativistic electron beams with truly compact and portable high repetition rate laser systems.

We thank Yuan Tay and Kiyong Kim for technical discussions regarding the gas jet, and Chenlong Miao and Dan Gordon for assistance with the PIC simulations. This work was supported by the Defense Threat Reduction Agency and the US Department of Energy.